\documentclass[12pt,aps,amsmath,latexsym,amsfonts]{JHEP3}
\usepackage[all]{xy}
\title{Stealth Acceleration and Modified Gravity}
\author{Christos Charmousis$^{\it 1}$, Ruth Gregory$^{\it{2}}$, 
and Antonio Padilla$^{\it {3}}$\\ 
{\it 1)} LPT, Universit\'e de Paris-Sud, 
B\^at. 210, 91405 Orsay CEDEX, France\\ 
{\it 2)} Centre for Particle Theory, Durham University,
South Road, Durham, DH1 3LE\\
{\it 3)} School of Physics \& Astronomy,
University Park,
University of Nottingham,
Nottingham NG7 2RD }
\date{\today}
\abstract{We show how to construct consistent braneworld models
which exhibit late time acceleration. Unlike {\it self-acceleration},
which has a de Sitter vacuum state, our models have the standard
Minkowski vacuum and accelerate only in the presence of matter, which
we dub ``{\it stealth}-acceleration''. We use an effective
action for the brane which includes an induced gravity term, 
and allow for an asymmetric set-up. 
We study the linear stability of flat brane vacua  and
find the regions of parameter space where the set-up is stable. 
The 4-dimensional  graviton is only
quasi-localised in this set-up and as a result gravity is modified at
late times.  One of
the two regions is strongly coupled and the scalar mode is eaten up by
an extra symmetry that arises in this limit. Having filtered the
well-defined theories we then focus on their cosmology. When the
graviton is quasi-localised we find two main examples of acceleration.
In each case, we provide an illustrative model and compare it to
$\La$CDM. }
\preprint{DCPT-07/29\\LPT-ORSAY/07-36}
\keywords{braneworlds, cosmology, modified gravity}
\usepackage{epsfig}
\usepackage{amsfonts,amssymb,amsmath}
\usepackage{subfigure}

\newcommand{\be}{\begin{equation}}
\newcommand{\ee}{\end{equation}}
\newcommand{\beq}{\begin{equation}}
\newcommand{\eeq}{\end{equation}}
\newcommand{\ba}{\begin{eqnarray}}
\newcommand{\ea}{\end{eqnarray}}
\newcommand{\bea}{\begin{eqnarray}}
\newcommand{\eea}{\end{eqnarray}}
\newcommand{\bean}{\begin{eqnarray*}}
\newcommand{\eean}{\end{eqnarray*}}
\newcommand{\bml}{\begin{mathletters}}
\newcommand{\eml}{\end{mathletters}}

\def\ga{{\bf \gamma}}

\def\half{\textstyle{\frac{1}{2}}}

\def\ie{{\it i.e.,}}
\def\M{{\mathcal{M}}}

\newcommand{\ka}{\ensuremath{\kappa}}

\newcommand{\La}{\ensuremath{\Lambda}}

\newcommand{\mn}{\ensuremath{{\mu\nu}}}

\newcommand{\mc}{\ensuremath{\mathcal}}
\newcommand{\del}{\ensuremath{\partial}}

\newcommand{\m}{\mathcal}

\renewcommand{\tt}{\textrm}

\begin{document}

\section{Introduction}

Recent observations of high red-shift supernovae suggest that our 
universe has entered a period of acceleration~\cite{champagnesupernova}. 
Combining these observations with the microwave background data, \cite{WMAP},
and the results of large scale structure surveys \cite{2DF}, we deduce
that the universe is approximately spatially flat, and that the
observed acceleration is consistent with standard $4D$ cosmology 
if we assume that roughly $70 \%$ of the actual energy content of the 
universe comes from a small positive cosmological constant, 
$\La/8\pi G \sim 10^{-12}$ (eV)${}^4$ \cite{concord}. 
Of the residual matter content,
only about $4 \%$ corresponds to baryonic matter, the remaining component 
corresponding to an also as yet unobserved ``dark matter sector". 
Although this simple energy pie chart fits the data extremely well, 
effective field theory methods have so far failed to give 
a satisfactory explanation for such a small value of the cosmological 
constant. Naive arguments
predict a cosmological constant of the order of $m_\textrm{pl}^4$, 
i.e.\ $10^{120}$ times larger than the ``observed'' value,
which fine tuning problem has inspired a search for 
alternative explanations for the observed acceleration.

An interesting alternative is that supernovae data are actually 
indicating the presence of a new physical scale where novel 
gravitational physics kicks in. The scale in question is a 
classical energy scale corresponding to the current Hubble curvature 
radius, $H_0 \sim 10^{-34}$ (eV). 
The braneworld scenario \cite{bw,RS} has been a natural breeding 
ground for these ideas (see, for example, 
\cite{multigravity,DGP,GRS,asymm}). Typically, gravity becomes 
weaker in the  far infra-red due to gravitational leakage into the 
extra dimensions. In some cases, this modification of gravity at 
large distances leads directly to exponential acceleration at 
late times, even when there is no effective cosmological 
constant~\cite{DGP,asymm,sa2}. These types of solutions are 
often referred to as {\it self-accelerating}, since their vacuum 
state corresponds to  a de Sitter brane. However, there is 
by now good evidence that self-accelerating solutions lead 
to problems with perturbative ghosts~\cite{DGPghosts}.

Modifying Einstein's General Relativity in a consistent 
manner in the infra-red has proven to be a very difficult task. 
There are three main reasons for this: theoretical consistency such as 
stability or strong coupling; experimental bounds mostly coming from 
the solar system; and naturalness, i.e.\ a restricted number of 
additional fields and parameters for the modified theory. The 
first problem, for the time-being at least, has been the most 
stringent one. Indeed, it is common for the vacua of such modified theories 
to suffer from  classical and/or quantum instabilities, at least at 
the level of linear perturbations. Most candidate braneworld 
models, \cite{multigravity,DGP,GRS} have been shown to suffer 
from such instabilities or strong coupling or both, 
\cite{DGPghosts,strong}. Generically, a ghost mode appears in the 
perturbative spectrum of the theory at the scale where gravity is 
modified, effectively driving the acceleration, much like a 
cosmological constant put in by hand. It has been argued that 
strong coupling, i.e.\ the breakdown of linear perturbation theory
\cite{dvali,screening}, may invalidate these claims,
bypassing also rather stringent solar system constraints. As 
a result however, in the absence of exact solutions
the gravity theory becomes unpredictable at the strong coupling 
scale and the loss of linear perturbation theory is 
perhaps a bigger problem to overcome.

The situation is not better settled for  four dimensional theories  that
modify the Einstein-Hilbert action directly. Here, early attempts go
back to Brans and Dicke \cite{BD}, who considered a massless
scalar-tensor theory with only one additional constant degree of
freedom, the kinetic coupling $\omega_{BD}$. Such a theory (which for
example goes through the naturality condition) has been ruled out by
solar system time delay experiments pushing $\omega_{BD}\gtrsim 40000$
coming from the Cassini satellite~\cite{cassini}. Non-minimal
scalar-tensor theories with some potential and hence varying $\omega$
can still be consistent with PPN experiments if we suppose that the
scalar field varies between solar system and cosmological
scales. Indeed, we remind the reader that the relevant cosmological
scale $H_0^{-1}$ and the size of the solar system give a dimensionless
quotient of $(1AU H_0)\sim 10^{-15}$ whereas the typical PPN parameter
$\gamma$ is 1 with error $10^{-5}$. Vector-tensor theories have also
been shown to be plagued by instabilities (the exception being the
case studied by Jacobson~\cite{jacobson} where the vector has a fixed
norm, and Lorentz-invariance is broken at the level of the
action{\footnote{At the level of cosmological solutions this is of
course true since there is no timelike Killing vector; the
difference there being that the symmetry is not broken at the
level of the action.}}). Adding higher order combinations of the
Riemann curvature tensor also generically leads to the appearance of
ghosts except if one considers functionals of the Ricci scalar
curvature $F(R)$ \cite{f(R)}. Then the theory in question can be
translated via a  conformal transformation to a non-minimal
scalar-tensor theory. Solar system constraints are again problematic
(though see \cite{Hu} for recent developments)
and often vacuum spacetimes such as Minkowski  
are not even solutions to the field equations (although perhaps
this is not such a big problem in the cosmological setting).

In this paper we present an alternative general class of ghost-free
braneworld models which possess accelerating solutions. 
However, unlike the usual self-accelerating braneworld solutions, 
and indeed unlike $\Lambda$CDM, our models do not necessarily lead 
to de Sitter cosmologies at late times. Indeed, the final fate of the
universe may not even be an accelerating cosmology! 
Also, unlike the usual models in which even the vacuum brane 
accelerates, in our models 
the vacuum state is a stable Minkowski brane,
and the Friedmann equation actually implies that 
while the Minkowksi brane is the vacuum state for an empty brane, once
one has an ``ordinary'' cosmological fluid, i.e.\ $p = w\rho$ with $w\geq0$,
then the brane {\it can} enter an era of acceleration, even though
there is no cosmological constant or dark energy present. 
Depending on the specific parameters in the model,
this acceleration can either persist, or the universe at some stage exits
from accelerated expansion into a stiff-matter cosmology. In either case,
the expansion is power law, rather than exponential.

Clearly an accelerating cosmology is of no interest if it
is hampered by the usual consistency problems, therefore we perform
a general scan of possible braneworld solutions, filtering out those
solutions that we suspect will contain problems with ghosts (such as
self-accelerating vacua or regions where we prove the appearence of ghosts). 
We also further restrict our parameter space by requiring that 
the bout of acceleration  is preceded by a
sufficient period of standard $4D$ cosmology, in order to reproduce
standard early universe cosmology, such as nucleosynthesis, the
development of structure etc., so that the standard cosmological
picture is retained. Note therefore, that these braneworld
models are not suited for explaining primordial inflation. In fact, the
basic feature of these models is that it is {\it only} at late times in a
matter dominated universe that this acceleration can kick in. Indeed,
if the parameters of the model were tuned so that the modified gravity 
scale were to become important during the radiation era, then the
universe would not accelerate, but merely `coast' \cite{coast}, before
returning to an expanding, decelerating, cosmology. Therefore, our
models explain quite naturally why it is only at scales comparable to 
the current Hubble time (or indeed longer) that acceleration can
occur.

The rest of this paper is organised as follows. In the next section we
describe our set-up involving a single brane embedded in some
five-dimensional bulk spacetime. This will be a combination of the
DGP~\cite{DGP}, and asymmetric brane~\cite{asymm} models. In other
words, we consider an asymmetric set-up, with an induced curvature
term on the brane. We will present the equations of motion, and derive
the vacuum states  corresponding to Minkowski branes. In section
\ref{pert} we consider vacuum perturbations, focusing on the
scalar radion mode. We will derive the effective action for this mode,
in order to filter out those solutions that contain a perturbative
ghost. In section~\ref{cosl} we introduce a cosmological fluid  and
derive the cosmological solutions, establishing the conditions
required for a consistent cosmology evolving towards a Minkowski
vacuum as the energy density decreases. The consistent solutions are
then analysed in greater detail in section \ref{analysis}, with
special emphasis on the conditions for cosmic acceleration. We analyse 
two main models in detail, showing how they give rise to cosmic
acceleration, and comparing these models with $\Lambda$CDM. 
Finally, we conclude in section  \ref{discussion}.

\section{Asymmetric braneworlds: formalism and set-up} \label{setup}

Consider a single 3-brane, $\Sigma$, embedded in between two bulk
five-dimensional spacetimes $\M_i$, where $i=1,2$. The brane can be
thought of as the common boundary, $\Sigma=\partial\M_1=\partial\M_2$
of these manifolds. Each spacetime $\M_i$ generically has a  five
dimensional Planck scale given by $M_i$, and a negative (or zero)
cosmological constant given by $\La_i=-6k_i^2$. In general we will not
consider $\mathbb{Z}_2$ symmetry and we will allow for the
cosmological constants, and even the fundamental mass scales to differ
on either side of the brane~\cite{asymm}.  Allowing for $\La_1 \neq
\La_2$ is familar enough in domain wall scenarios \cite{DW}. Here we are also
allowing for $M_1 \neq M_2$. This is not so familiar, but could arise
in a number of ways. Suppose, for example, that this scenario is
derived from a fundamental higher dimensional theory.  This theory
could contain a dilaton field that  is stabilised in different
fundamental vacua on either side of $\Sigma$. From the point of view
of a $5D$ effective description,  the $5D$ Planck scales would then
differ accordingly. Indeed naive expectations from string theory point
towards this asymmetric scenario as opposed to a symmetric one.
Different effective Planck scales can also appear on either side of a
domain wall that is bound to a five-dimensional  braneworld~\cite{nested}. 

The brane itself has some vacuum energy, or tension. This will
ultimately be fine-tuned against the bulk parameters, $\La_i$ and
$M_i$,  in order to admit a Minkowski vacuum solution. In other words,
there will be {\it no effective cosmological constant on the
brane}. As in the original DGP model~\cite{DGP},  we will also allow
for some intrinsic curvature to be induced on the brane. Such terms
are rather natural and can be induced by matter loop
corrections~\cite{loops},  finite width effects~\cite{width} or even
classically from higher dimensional modifications of General 
Relativity~\cite{z}.

Our set-up is therefore described by  the general 5
dimensional action,
\begin{equation}
\label{act}
S=S_\textrm{bulk} +S_\textrm{brane}
\end{equation}
The bulk contribution to the action is described by
\begin{equation}
S_\textrm{bulk}=\sum_{i=1,2} M_i^3\int_{\mathcal{M}_i}
\sqrt{-g}(R-2\Lambda_i)+2M_i^3\int_{\partial\mathcal{M}_i}
\sqrt{-\gamma} K^{(i)} \label{bulk}
\end{equation}
where $g_{ab}$ is the bulk metric with corresponding Ricci tensor,
$R$. 
The metric induced on the brane ($\partial \m{M}_i$) is  given by
\be
\gamma_{ab}=g_{ab}-n_an_b
\ee
where $n^a$ is the unit normal to $\partial\mathcal{M}_i$ in
$\mathcal{M}_i$ pointing {\it out} of $\mathcal{M}_i$. Of course,
continuity of the metric at the brane requires that $\gamma_{ab}$ is
the same, whether it is calculated from the left, or from the right of
the brane. In contrast, the extrinsic curvature  of the brane can jump
from right to left. 
In $\partial \m{M}_i$,  it is defined as
\begin{equation} K^{(i)}_{ab}=\gamma^c_a
\gamma^d_b \nabla_{(c} n_{d)} \label{extrinsic}
\end{equation}
Its trace  appears in the Gibbons-Hawking boundary term in (\ref{bulk}).

The  brane contribution to  the action, meanwhile, is described by
\begin{equation}
S_\textrm{brane}=\int_\textrm{brane}\sqrt{-\gamma}(m_{pl}^2
\mathcal{R}-\sigma+\m{L}_\tt{matter}) \label{brane} \end{equation}
where $\sigma$ is the brane tension, and $\m{L}_\tt{matter}$ includes
any matter excitations. We have also included the  induced intrinsic
curvature  term, $\m{R}$, weighted by a $4D$  mass scale,
$m_\tt{pl}$. Note that we have taken $m_\tt{pl}^2>0$, as in the
original DGP model. There are two reasons for this. Firstly, when it
comes to studying the cosmological solutions, we would like this term
to dominate the cosmology at early times, in order to reproduce the
standard $4D$  cosmology, as discussed in the introduction. Secondly,
allowing $m_\tt{pl}^2<0$ could result in vacuum perturbations
containing a spin-2  ghost~\cite{ghbig}.

The equations of motion in the bulk region, $\m{M}_i$,  are just the
 Einstein equations, with the appropriate cosmological constant, $\La_i$.
\begin{equation}
E_{ab}= R_{ab}-\frac{1}{2} R g_{ab}+\Lambda_i g_{ab}=0
\label{bulkeom}
\end{equation}
The equations of motion on the brane are described by the Israel
junction conditions, and can be obtained by varying the action
(\ref{act}), with respect to the brane metric, $\gamma_{ab}$.  This gives
\begin{equation} \Theta_{ab}=2\left \langle M^3 (K_{ab}-K
\gamma_{ab}) \right \rangle+m_{pl}^2\left( \mathcal{R}_{ab}-\frac{1}{2}
\mathcal{R}\gamma_{ab}\right)+\frac{\sigma}{2}\gamma_{ab}=\frac{1}{2} T_{ab}
\label{braneeom}
\end{equation}
where $T_{ab}=-\frac{2}{\sqrt{-\ga}}\frac{\partial
  \sqrt{-\ga}\m{L}_\tt{matter}}{\partial \ga^{ab}}$. The angled
brackets denote an averaged quantity at the brane. More precisely, for
some quantity $Q_i$ defined on the brane in $\partial \m{M}_i$, we
define the average 
\be
\langle Q \rangle= \frac{Q_1+Q_2}{2}\; .
\ee
Later on we will
also make use of the difference,  $\Delta Q =Q_1-Q_2$.
Note that the Israel equations here do not use the familiar
``difference'', because we have defined the unit normal as pointing
out of $\mathcal{M}_i$ on each side, i.e.\ the approach 
is that of the brane as a boundary. Israel's equations on the 
other hand were derived for thin
shells in GR, i.e.\ where the brane is a physical, very thin, object, and
the normal is thus continuous, pointing `out' on one side of the wall,
and `in' on the other.

We will now derive the vacuum solutions to the equations of motion
(\ref{bulkeom}) and (\ref{braneeom}). This corresponds to the case
where there are no matter excitations, and so, $T_{ab}=0$.  In each
region of the bulk, we introduce coordinates $x^a=(x^\mu, y)$, with
the brane located at $y=0$. We will not be interested in de Sitter
solutions, since these will only arise through an excess in vacuum
energy, or through ``self-acceleration''. The former offers no
alternative to $\La$CDM, whereas the latter is expected to suffer from
a generic ghost-like instability~\cite{DGPghosts}. Therefore, we 
seek solutions of the  form
\be
ds^2=\bar g_{ab}dx^adx^b=a^2(y)\eta_{ab} dx^a dx^b \label{backg}
\ee
Inserting this into the bulk equations of motion (\ref{bulkeom}) gives
\be
\left( \frac{a'}{a}\right)^2=k^2a^2, \qquad \frac{a''}{a}=2k^2a^2 
\label{bulkaeq}
\ee
where ``prime'' denotes differentiation with respect to $y$. Note that
we have dropped the asymmetry index $i$ for brevity. Equations
(\ref{bulkaeq}) have  solution
\be
a(y)=\frac{1}{1-\theta ky} \label{a}
\ee
where $\theta=\pm 1$. Note that each region of the bulk corresponds to $0<y<y_\tt{max}$, where
\be
y_\tt{max}=\begin{cases} 1/k & \tt{for $\theta=1$} \\
\infty & \tt{for $\theta=-1$} \end{cases}
\ee
For $k \neq 0$, this means that when $\theta=1$ we are keeping the adS
boundary (growing warp factor) whereas when $\theta=-1$ we are keeping
the adS horizon (decaying warp factor). For $k=0$, we simply have a
Minkowski bulk, in the usual coordinates, and the sign of $\theta$ is
irrelevant.

The boundary conditions at the brane (\ref{braneeom}) lead to a 
finely tuned brane tension
\be
\sigma=-12\langle M^3 \theta k\rangle \label{tension}
\ee
This fine tuning guarantees that there is no cosmological constant on
the brane, and is equivalent to the (asymmetric) Randall-Sundrum
fine-tuning for $\theta_1k_1<0$ and  $\theta_2k_2<0$~\cite{RS}.

\section{Linearised vacuum perturbations and asymptotic stability} \label{pert}
We shall now consider linearised perturbations, $h_{ab}$ about our
 background solutions (\ref{backg}) and (\ref{a}), so that
\be
ds^2=a^2(y)\left[\eta_{ab}+h_{ab}(x, y)\right] dx^a dx^b
\ee
In the unperturbed spacetime, the gauge was fixed in both $\m{M}_1$
and $\m{M}_2$ so that the brane was at $y=0$. However, a general
perturbation of the system must also allow the brane position to
flutter. In $\m{M}_i$, the brane will be located at
\be
y=f_i(x^\mu)
\ee
Of course, these expressions contain some gauge-dependence due to
invariance under the following diffeomorphism  transformations
\be
y \to y+\eta(x, y), \qquad x^\mu \to x^\mu+\zeta^\mu(x, y) \label{trans}
\ee
Now, it is convenient and physically relevant to decompose these
transformations in terms of the $4D$ diffeomorphism group. This gives
$\zeta^\mu=\xi^\mu+\del^\mu \xi$, where $\xi^\mu$ is a Lorentz-gauge
vector satisfying $\del_\mu \xi^\mu=0$. We do likewise for the  perturbation: 
\bea
h_\mn &=& h_\mn^\tt{TT}+2\del_{(\mu} F_{\nu)}+2\del_\mu \del_\nu 
E+2A\eta_\mn \label{mn}\\
h_{\mu y} &=& B_\mu+\del_\mu B \\
h_{yy} &=&2\phi \eea Again, $F_\mu$ and $B_\mu$ are Lorentz-gauge
vectors, whereas $h_\mn^\tt{TT}$ is a transverse-tracefree tensor,
$\del_\mu h^\tt{TT}{}^\mu{}_\nu=h^\tt{TT}{}^\mu{}_\mu=0$. Note that
greek indices are raised and lowered using $\eta_\mn$. Under the
gauge transformations (\ref{trans}), the various components of the
perturbation transform as follows \be h_\mn^\tt{TT} \to
h_\mn^\tt{TT} \ee \be B_\mu \to B_\mu-\xi_\mu', \qquad F_\mu \to
F_\mu-\xi_\mu \ee
\be  \phi \to \phi-\frac{(\eta a)'}{a}, \qquad B
\to B-\xi'-\eta, \qquad E \to E-\xi, \qquad A \to A-\frac{a'}{a}
\eta \ee We immediately see that the tensor component,
$h_\mn^\tt{TT}$ is gauge invariant. We can also construct the
following vector and scalar gauge-invariants in the bulk \be
X_\mu=B_\mu-F_\mu', \qquad X=A-\frac{a'}{a}(B-E'), \qquad
Y=\phi-\frac{[a(B-E')]'}{a} \ee We will now consider vacuum
fluctuations, so that the bulk equations of motion and the Israel 
junction condition are given by $\delta E_{ab}=\delta
\Theta_{ab}=0$. At this point we assume that the tensors,
vectors and scalars do not mix with one another, so that their
equations of motion can be taken independently. We will come back and 
check the validity of this hypothesis against our results later on. 
Focusing on the
scalars, we find that the bulk equations of motion give $\delta
E_{ab}^\tt{scalar}=0$, where \be \delta
E^\tt{scalar}_{\mu\nu}=\left[-\del_\mu \del_\nu +\eta_{\mu\nu}\del^2
\right](2X+Y)+3\eta_{\mu\nu} \left[\partial_y+3 \frac{a'}{a}
\right]\left(X'-\frac{a'}{a}Y \right) \ee \be \delta
E^\tt{scalar}_{\mu y} = -3 \del_\mu \left(X'-\frac{a'}{a}Y \right)
\ee \be \delta E^\tt{scalar}_{yy} = 3 \del^2 X+12\frac{a'}{a}
\left(X'-\frac{a'}{a}Y \right) \ee These equations are easily solved
on each side of the brane (we remind the reader that we have dropped
the asymmetry index $i$) to give 
\be X=\frac{U(x)}{a^2}, \qquad Y=-\frac{2U(x)}{a^2} 
\ee
where $\del^2 U=0$.  We can identify $U(x)$ as the bulk radion mode.
It represents two degrees of freedom, $U_i(x)$, $i=1,2$,  
one on each side of the
brane. In addition to this, we have the brane bending degrees of freedom,
given by $f_1$ and $f_2$. Given that we have two boundary conditions
at the brane, namely continuity of the metric and the Israel
equations, we expect the number of physical scalar degrees of
freedom to be at most two, as we shall now demonstrate.

To impose the boundary conditions at the brane it is convenient to
work in brane-GN gauge. In this gauge, we have  $B=\phi=0$ and the
brane is fixed at $y=0$. It follows  that
\begin{equation}
E=W(x)+V(x)Q(y)-U(x)Q(y)^2, \qquad A=\frac{U(x)}{a^2}-\frac{a'}{a^2}
\left[V(x)-2U(x)Q(y)\right]
\end{equation}
where
\begin{equation}
Q(y)=\int_0^y \frac{dy_1}{a(y_1)}=\frac{1}{2\theta k}\left[1-a^{-2}(y) \right]
\end{equation}
The brane bending degrees of freedom are now encoded in the fields,
$V_1$ and $V_2$. In contrast, $W_1$ and $W_2$ merely reflect the
freedom to choose the gauge along the brane, and as such, can be
consistently set to zero. To see this, we evaluate our solution at
$y=0$ to derive the scalar part of the brane  metric (\ref{mn})
\be
h_\mn^\tt{scalar}(x, 0)=2 \del_\mu \del_\nu W+2 [U-\theta kV] \eta_{\mu\nu}
\ee
The pure gauge part, $2 \del_\mu \del_\nu W$, and the remainder $2
[U-\theta kV] \eta_{\mu\nu}$ need to be well defined {\it
  independently} of one another. This means that $\Delta W=0$, and it
is easy to to see that we can continously gauge away $W$ on both sides
of the brane, so that  $W=0$.

The continuity condition at the brane and the Israel equations now require that
\be
\Delta[U-\theta kV]=\left\langle M^3V +m_{pl}^2  [U-\theta
  kV]\right\rangle=0 
\label{bc4UV}
\ee
Therefore, as predicted, we are left with at most two scalar degrees
of freedom as perceived by a 4 dimensional observer. Actually, it is
easy to see why this is so if one introduces regulator branes at (say)
$y=y_*< y_\tt{max}$. The remaining degrees of freedom essentially
correspond to the fluctuations in the proper distance in between
branes. Actually, the introduction of regulator branes is particularly
useful when trying to impose boundary conditions in the bulk, and in
terms of calculating a $4D$ effective action. We shall now develop
this in  more detail.

In the background, the extrinsic curvature, $K_{\mu\nu}^*$, of the 
regulator brane satisfies the following
\be
K_{\mu\nu}^*+\frac{a'(y_*)}{a^2(y_*)}\gamma^*_{\mu\nu}=0 \label{regbc}
\ee
where  $\gamma^*_{\mu\nu}$ is the induced metric on the regulator
brane. We require that this equation also holds in the perturbed
scenario. This time, it is easiest to work in GN coordinates relative
to the {\it regulator brane} (we shall call these regulator-GN
coordinates). The regulator brane is now fixed at $y=y_*$, and $B=\phi
=0$. Again, it follows  that
\begin{equation}
E=W_*(x)+V_*(x)Q_*(y)-U(x)Q_*(y)^2, \qquad
A=\frac{U}{a^2}-\frac{a'}{a^2}
\left[V_*(x)-2U(x)Q_*(y)\right]
\end{equation}
where
\be
Q_*(y)=\int_{y_*}^y \frac{dy_1}{a(y_1)}=\frac{1}{2\theta k}
\left[a^{-2}(y_*)-a^{-2}(y) \right]
\ee
As before we can set $W_*=0$. Applying the boundary condition
(\ref{regbc}) 
we deduce that $V_*=0$.

It is very important to realise that the brane-GN coordinates  and the
regulator-GN coordinates are not necessarily the same. In particular,
in brane-GN coordinates the true brane is at $y=0$, but the regulator
brane might {\it not} be at $ y=y_*$. Of course the two sets of
coordinates must be related by a gauge transformation. Starting in
regulator-GN gauge, we transform to brane-GN gauge with the following
coordinate  change
\be
x^\mu \to x^\mu+D^\mu \xi, \qquad y\ \to y+\eta
\ee
where
\be
\xi=U\left[Q(y)^2-Q_*(y)^2 \right]-VQ(y), \qquad 
\eta=\frac{V}{a}-\frac{2U}{a}\left[Q(y)-Q_*(y)\right]
\ee
From this we can deduce that in brane-GN coordinates the regulator 
brane is at $y=y_*+\eta(x, y_*)$ where
\be
 \eta(x, y_*)=\frac{V-2UQ(y_*)}{a(y_*)} \label{bending}
\ee
We would now like to calculate the proper distance between the
branes. To do this carefully, it is best to work in a gauge in which
both the true brane {\it and} the regulator brane are fixed. Clearly
neither brane-GN nor regulator-GN gauge will do. Instead,  we start
off in brane-GN coordinates,  and transform to fixed brane coordinates
by  letting
\be
x^\mu \to x^\mu, \qquad y \to y+\epsilon(x, y),
\ee
so that
\be
\phi \to -\frac{(\epsilon a)'}{a}, \qquad B \to -\epsilon, \qquad E\to
E, 
\qquad A \to A-\frac{a'}{a}\epsilon 
\label{fixed}
\ee
If we set $\epsilon(x, 0)=0$ and $\epsilon(x, y_*)=-\eta(x, y_*)$,
both branes will be fixed,  at $y=0$ and $y= y_*$ respectively. The
proper distance between them is given  by
\be
 z_*=\int_0^{y_*} dy \sqrt{g_{yy}}=\bar z_*+\delta z_*
\ee
where 
\be
\bar z_*=-\frac{1}{\theta k}\ln(1-\theta k y_*)
\ee
is the proper distance between branes in the background, and
\be
\delta z_*=V-2Q(y_*) U
\ee
is the fluctuation. Fixed wall gauge is also useful for calcluating
the effective action. If we assume that the boundary conditions hold
at the true brane and at both regulator branes, the effective action
(to quadratic order) is simply given by the following bulk  integral
\be
S_\textrm{eff}=\half \int_\textrm{bulk} \sqrt{-\bar g}M^3 
\delta g^{ab}\delta E_{ab}
\ee
Choosing fixed brane gauge in both the left hand bulk and the right
hand bulk, we  find
\be
S_\textrm{eff}= -6\int d^4x  {\Big\langle} M^3 \delta z_*\del^2
 U(x) {\Big\rangle} \label{seff1}
\ee
where $U$ can be related to the fields $\delta z_*$ using the boundary
conditions at the brane (\ref{bc4UV}). Now consider what happens, as
we remove the regulators by taking the limit $y \to y_\tt{max}$, or
equivalently $\bar z_* \to \infty$.  For $U \neq 0$, 
\be
\delta z_* \to \begin{cases} -A/k & \textrm{whenever $\theta k>0$} \\
 \infty &  \textrm{whenever $\theta k\leq 0$} \end{cases} \label{Q}
\ee
Therefore, if $\theta k \leq 0$, the fluctuation in the proper
distance between branes diverges. This reflects the fact that the
gauge invariant mode, $U$, is non-normalisable. Since we must have
normalisable boundary conditions in order to obtain a {\it local} $4D$
effective theory on the brane, we require that $U=0$ whenever $\theta
k\leq 0$. The result is that the scalar degree of freedom coming from
a given bulk region only survives the single brane limit when the
corresponding regulator is taken to the AdS boundary as opposed to the
AdS horizon. This makes sense for the following reason. If an observer
on the brane shines a light ray  into a bulk region towards the AdS
boundary, he/she sees it reflected back from the boundary in a finite
proper time. However, if the ray is shone into the bulk towards the
AdS horizon, it will never be reflected back. One is  therefore able
to detect fluctuations in the proper distance between the brane and
the AdS boundary, but not the brane and the AdS  horizon.

After removing the regulators, and imposing the boundary conditions
(\ref{bc4UV}), we find that there are no normalisable scalar degrees
of freedom left if $\theta_1k_1, \theta_2k_2<0$.  This does not come as a 
surprise since in this case the bulk includes the AdS horizon on both sides
of the brane, and so one would not expect there to be any normalisable
radion degrees of freedom for the reasons outlined in the previous
paragraph. In all other cases, the effective action (\ref{seff1}) is
given  by
\be
S_\tt{eff}=6\left[m_\tt{pl}^2-\left \langle \frac{M^3 \theta}{k}
\right \rangle\right] \int d^4 x (\del A)^2 \label{seff2}
\ee
where the field $A=A(x, 0)=U-\theta k V$ measures conformal rescalings
of the metric on the brane. It
represents the only scalar degree of freedom that survives, since it
is the only scalar in the induced metric that cannot be locally gauged
away. Physically, however, we might have expected there to be two
radion degrees of freedom since both sides of the bulk included the
AdS boundary. Careful inspection however, shows that this is not the
case. Indeed reintroducing the regulator branes, we can split the  two
degrees of freedom into a centre of mass motion of the brane and a
relative motion \cite{radion}. The latter is shown to drop out as we push the
regulators out to the boundary. Furthermore, note that even the centre
of mass motion drops out if we have a zero cosmological constant on
either side of the brane. This is because in flat spacetime we have an
extra translation Killing vector in our coordinate chart (with respect
to adS) that permits us to consistently gauge away the remaining
scalar  mode.

Before we start to derive conclusions  about when there is a scalar
ghost in the spectrum of perturbations, let us pause for a moment to
discuss the reliability of these results.  In deriving the scalar
effective action we assumed that the scalars did not mix with the
vectors or tensors. Now, we need not worry about vectors, because the
vector contribution can always be locally gauged away on the
brane. The same cannot be said for the tensors, which are gauge
invariant, so there is a danger that they could mix with the scalars,
as was the case for the self-accelerating DGP
solution~\cite{DGPghosts}. To see whether this can happen here, we
write the scalar gauge-invariant piece in Gaussian-Normal gauge as  follows
\be
h_\mn^{(U)}=-\frac{1}{2 k^2 a^4}\del_\mu \del_\nu U \label{rad}
\ee
Because $\del^2 U=0$, we can view $h_\mn^{(U)}$ as a massless
transverse-tracefree tensor. Therefore, we can trust our analysis
provided there are no massless modes in the tensor sector. This is
guaranteed if the background bulk has infinite volume. The volume of
the background is {\it finite} if and only if $\theta_1k_1$ and
$\theta_2k_2$ are both strictly negative.  This corresponds to a
generalised Randall-Sundrum scenario where the warp factor decays into
the bulk. Then indeed there is a tensor zero mode, and we have mixing
between tensors and scalars at zero mass. Although our analysis may
therefore be unreliable in this instance, it is well known that in
this case there is actually no radion degree of freedom. In fact, the
mixing actually  ensures that the graviton propagator has the correct
$4D$ tensor structure~\cite{GT} as in the original RS scenario. It
also follows that there can be no infra-red modification of gravity
since the graviton zero-mode always guarantees Einstein like behaviour
at large distances and at late times. This will not be so interesting
from an infra-red  cosmological perspective, so from now on, we will
drop the case  $\theta_1k_1, ~\theta_2k_2<0$.

In all other cases, the bulk volume is infinite and there is {\it no}
normalisable zero-mode graviton in the spectrum and hence no
mixing. However, we do want a {\it quasi-localised} graviton
dominating up to some energy scale in the graviton spectrum as one
gets for DGP~\cite{DGP} or GRS~\cite{GRS}. This is guaranteed by the
induced curvature on the brane entering with the ``correct'' sign
($m_\tt{pl}^2>0$), as we will demonstrate in the next
section. Whenever the quasi-localised zero mode dominates, the higher
mass modes and the radion mix giving some type of well-defined
generalised scalar-tensor gravity. At large distances, however, the
quasilocalised nature of the graviton disappears giving a continuum of
massive modes with no mass gap.  Then
the radion mode (\ref{rad}) no longer mixes and is dangerous in the
sense that it can be a ghost. Whenever $k_1k_2=0$ we have already
discussed how the radion mode disappears. However, if $k_1k_2 \neq
0$, the radion is present, and we see from (\ref{seff2}) that it is
{\it not a ghost},  provided 
\be
\chi=m_\tt{pl}^2-\left \langle \frac{M^3 \theta}{k}\right \rangle \leq 0
\ee
It is interesting to note the competition between the bulk term in
$\chi$, and the induced gravity term: the ``correct" sign for the  DGP
term ($m_\tt{pl}^2>0$)  {\it always} contributes to a ghost-like
radion (as does a localised warp factor). This behaviour is caused by
the so-called conformal ghost that appears in $4D$ Einstein gravity
for perturbations about Minkowksi space (see for
example~\cite{confghost}). In that case the scalar mode is harmless
since it provides the correct tensor structure in the propagator
mixing with the zero-tensor mode. Strictly speaking each individual
mode doesn't exist independently: it is only their linear combination
which has physical meaning. However, in our case the absence of a
graviton zero-mode ensures that there is no mixing between modes and
so a radion ghost always signals a vacuum  instability.

Of particular interest are the limits $\chi \to 0$, and $\chi \to
\infty$. For small/large $\chi$, the radion  couples very
strongly/weakly to the trace of the energy-momentum tensor, since
schematically we  have,
\be
-\chi \partial^2 A \sim T
\ee
In the $\chi \to \infty$ limit, the radion completely decouples. This
corresponds to the case where $k_1k_2=0$ so at least one side of the
bulk is Minkowski. The radion decouples in this case because
it costs no effort to translate the brane toward the Minkowski 
side, and hence no brane bending can be detected.

The $\chi \to 0$ limit might also be refered to as the {\it conformal}
limit, since the brane can only support conformal matter sources
($T=0$) to linear order in perturbation theory. If we introduce some
non-conformal matter, the linearised theory breaks down because of
strong coupling, and the geometry responds non-linearly. This
behaviour reflects the onset of a new symmetry. It is reminiscent of
the partially massless limit ($m^2=2H^2$) of a  massive graviton
propagating in de Sitter space~\cite{DesWal}. In that theory an extra
symmetry kicks in that eliminates the scalar degree of freedom. In our
case, the linearised field equations, $\delta E_{ab}=0$ and $\delta
\Theta_{ab}=\half T_{ab}$, become invariant under the transformation
$h_{ab} \to h_{ab}+ h^{(f)}_{ab}$  where
\be
h_{\mu\nu}^{(f)}=(1-a^{-2})\partial_\mu \partial_\nu f-2k^2f 
\eta_{\mu\nu}, \qquad h^{(f)}_{\mu y}= h^{(f)}_{yy}=0
\ee
This is pure gauge in the bulk, but not on the brane. The
transformation therefore encodes an extra symmetry, beyond the usual
diffeomorphisms, that eliminates the radion degree of freedom when  $\chi=0$. 

In summary, if the bulk has infinite volume, we have a theory without
a normalisable tensor zero-mode, for which the radion will dominate at
large distances. In certain limits, the radion either decouples ($\chi
\to \infty$), or is eliminated by a new symmetry ($\chi \to
0$). Otherwise, the radion is present, and will render the vacuum
unstable on large scales unless  $\chi \leq 0$.

\section{Cosmological solutions} \label{cosl}

Let us now consider what happens when we introduce a cosmological
fluid to the Minkowski braneworlds derived in section \ref{setup}. In
order to preserve homogeneity and isotropy on the brane, we must
assume that the bulk metric is a warped product of the  form \cite{cosmo}
\be
ds^2=\mathcal{A}^2(t, z)(-dt^2+dz^2)+\mathcal{B}^2(t, z)d{\bf x}_\kappa^2
\ee
where for $\ka=1, 0, -1$, $d{\bf x}_\kappa^2$ is the metric on the unit
3-sphere, plane, and hyperboloid respectively. It turns out that we
have enough symmetry to render  the bulk equations of motion
(\ref{bulkeom}) integrable, and a generalised form of Birkhoff's
theorem applies \cite{cosmo}. When $\La=\ka=0$, the 
bulk solution is just a portion
of Minkowski spacetime, whereas in all other cases the bulk is a
portion of a black hole spacetime with cosmological constant,
$\La=-6k^2$, and horizon geometry parametrised by  $\ka$:
\be
ds^2=-V(r)dt^2+\frac{dr^2}{V(r)}+r^2d{\bf x}_\kappa^2, \qquad
V(r)=k^2r^2+\kappa-\frac{\mu}{r^2} \label{schw}
\ee
In order to construct the brane, we glue a solution in $\mathcal{M}_1$
to a solution in  $\mathcal{M}_2$, with the brane forming the common
boundary. Let us describe this in more detail. Assume for the moment
that the bulk solution on both sides of the brane takes the form 
(\ref{schw}). In $\mathcal{M}_i$, the boundary, $\partial \mathcal{M}_i$, 
is given by the section $(t_i(\tau), r_i(\tau), {\bf x}^{\mu})$ of 
the bulk metric. The parameter $\tau$ is the proper time of an observer 
comoving with the boundary, so that
\be \label{unit}
-V_i(r_i)\dot{t}^2 + \frac{\dot{r}^2}{V_i(r_i)} = -1,
\ee
where overdot corresponds to differentiation with respect to
$\tau$. The {\it outward} pointing unit normal to $\partial \mathcal{M}_i$
is now given by
\be
n_a=\theta_i( \dot r_i(\tau), -\dot t_i(\tau), {\bf 0})
\ee
where $\theta_i= \pm 1$. For $\theta_i=1$, $\mathcal{M}_i$
corresponds to  $r_i(\tau) < r< \infty$, whereas for  $\theta_i=
-1$, $\mathcal{M}_i$ corresponds to  $0 \leqslant r < r_i(\tau)$. The
signs of $\theta$ are consistent with the analysis in the previous two
sections.

The induced metric on $\partial \mathcal{M}_i$ is that of a FRW universe,
\be \label{eq:frw}
d s^2 = -d \tau^2 + r_i(\tau)^2 d {\bf x}_\ka^2,
\ee
Since the brane coincides with both boundaries, the metric on the
brane is only well defined when $r_1(\tau)=r_2(\tau)=R(\tau)$. The
Hubble parameter on the brane is now defined
by $H=\dot R/R$ where $R=R(\tau)$ is the brane trajectory 
in the bulk spacetime.

The brane equations of motion are again given by the Israel equations
(\ref{braneeom}). In addition to the finely tuned tension, we
introduce a homogeneous and isotropic fluid on the brane, with energy
density, $\rho$, and pressure, $p$. We will assume that these satisfy
the strong energy condition, so that $\rho \geq 0$ and $\rho+3p \geq
0$. This guarantees that any cosmic acceleration that may occur is
entirely due to modified gravity. The Israel equations now give the following
\be
\frac{\sigma+\rho}{6}=m_{pl}^2\left(H^2+\frac{\kappa}{R^2} \right)
-2\left\langle M^3 \theta
\sqrt{H^2+\frac{V(R)}{R^2}} \right \rangle \label{fried}
\ee
where the tension, $\sigma$, is given by equation
(\ref{tension}). Note this formula is general and
holds for all values of parameters, even those involving Minkowski
backgrounds, for which we simply set $V(R)=0$ in (\ref{fried}). 

For simplicity, we will assume from now on that the bulk spacetime
is maximally symmetric by setting the mass term, $\mu=0$. One can confirm
that this term, which has the behaviour of dark radiation, 
does not play an important role at late times, when 
it is subdominant to the matter content on the brane. In
addition, since observations of the first acoustic peak in the Cosmic
Microwave Background demonstrate that the universe is very nearly
flat \cite{WMAP}, we set $\kappa=0$.   The modified Friedmann equation then
takes the compact form
\be
\rho=F\left(H^2\right) 
\label{rho}\ee
where
\be
F({H}^2)=6m_{pl}^2{H}^2-12\left\langle M^3 \theta
\left(\sqrt{{H}^2+k^2}-k \right)\right\rangle \label{F}
\ee
Now, as $H \to 0$ , it is easy to check that the energy density $\rho
\to 0$. This serves as a consistency check that the finely tuned
tension (\ref{tension}) guarantees that the Minkowski brane
corresponds to a possible vacuum solution.  There might be other
vacuum solutions with $H_0 > 0$, corresponding to
self-accelerating vacua with $F(H_0^2)=0$. These however are 
suspected to contain
perturbative ghosts~\cite{DGPghosts} (although we will not attempt to
show this explicitly here). We will return briefly to discuss how and
when such vacua arise presently. Also note
that switching off $M_i$ we get the usual Friedmann  equation,
and switching off $m_{pl}^2$ gives the usual Randall-Sundrum modified
Friedmann equation \cite{nucleo}.

Since we wish to study the possibility of cosmic acceleration in these
models, it is natural to examine the cosmic deceleration parameter,
$q=-\frac{\ddot R R}{~\dot R^2}$. Assuming a constant equation of
state, $p=w \rho$, it follows from (\ref{rho}) and conservation
of  energy
\be
\dot \rho=-3H(\rho+p) \label{energy}
\ee
 that the deceleration parameter is given by
\be
q=-1+\frac{3}{2}(1+w) \, \m{C}(H^2) 
\ee
Here the functional 
\be 
\m{C}(H^2) = \frac{F(H^2)}{H^2 F'(H^2)},
\ee 
will vary during the course of the cosmological evolution whereas for
GR we have simply $\m{C}=1$. It follows that the deceleration
parameter will also vary during the cosmological evolution, in
contrast to what happens in four dimensional Einstein gravity where it
is constant for constant $w$. For cosmic acceleration to occur,
obviously the deceleration parameter must become negative. For
ordinary forms of matter satisfying the strong energy condition
$1+3w \geq 0$, it is easy to check that acceleration can only be
achieved if $\m{C}(H^2)$ falls below  one.

It is useful to define an {\it effective} equation of state parameter 
for the universe, $\gamma_\tt{eff}$, which is related to the deceleration
parameter $q=\frac{1}{2}(1+3\gamma_\tt{eff})$,  from which we deduce  that
\be
1+\gamma_\tt{eff}=\m{C}(1+w) \label{C}
\ee
Again, as $\m{C}$ varies during the cosmological evolution, so 
must the effective equation of state. 
Note however, this should not be confused with a {\it dark energy}
component with a varying equation of state (see e.g.\ \cite{varyw}), 
in which there are two main components of the energy of the universe:
matter and the time-varying dark energy. Here, we assume
that there is only {\it one} dominant
cosmological fluid, and $\gamma_\tt{eff}$ provides a simple, effective
way of tracking the gravitational effect of that fluid.
It is clear that the cosmological behaviour is completely determined
by the functional, $F$, and its derivatives, with the combination
$\m{C}=F/F'H^2$ proving particularly important when asking questions
about cosmic acceleration. 
To emphasize the difference between taking
{\it dark energy}, in which we modify the matter in the Friedmann equation
to $H^2 = \m{F}(\rho)$, and {\it light gravity} in which we modify
the geometry: $\rho = F(H^2)$, consider the $\Lambda$CDM model. If
we regard this in its conventional dark energy way, we have the dark 
energy with a constant equation of state: $w=-1$. If, however, we
regard the cosmological constant as part of the gravitational sector,
then we obtain $ \gamma_\tt{eff}^{(\Lambda\tt{CDM})} = 
- \Omega_\Lambda H_0^2/H^2$ for a matter cosmology.

As the universe expands, it is clear from the strong energy condition,
and energy conservation (\ref{energy}), that the energy density of the
universe must decrease. As the energy density is diluted, we ought to
approach a vacuum state, which will either correspond to a Minkowski
brane, or a self-accelerating brane. We are interested in the case
where we approach the Minkowski vacuum. Now, in the absence of any
phase transitions, we expect $H$ to vary continously during the
cosmological expansion. It follows that our brane cosmology must be
consistent over a range $0 \leq H <  H_\tt{max}$.

To be consistent, our cosmology must adhere to certain rules. To begin
with, the strong energy condition requires that the energy density is 
non-negative. 
It only vanishes for the vacuum brane, which we have taken to be the
Minkowski solution. We therefore require that $F(H^2)>0$ over the
range $0<H<H_\tt{max}$ (see figure 1). It is easy to check that this
rule implies  that
\be
F'(0)=6\chi \geq 0 \label{F'0}
\ee  
The alternative scenario, where $F'(0)<0$, is also shown 
in figure \ref{sampleF}. 
Here there is a small region close to  $H_\epsilon>0$ for which 
$F(H_\epsilon)<0$ and therefore $\rho_\epsilon<0$. This is an
unphysical regime. The only way to get a physical vacuum is for  $F'$
to change sign, so that we have a {\it self-accelerating} vacuum  with
$H=H_0$. The Minkowski brane in this instance represents an {\it
isolated} vacuum which will never be approached, even as the energy
density is  diluted. 
\EPSFIGURE[left]{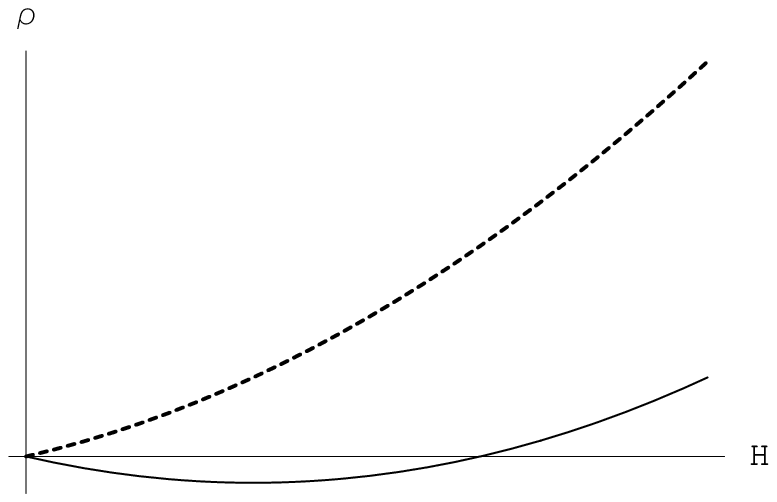}{An example of two possible 
functions $ \rho = F(H^2)$ where there is a Minkowski vacuum (the dotted 
line), and where the Minkowski vacuum is isolated (solid line). These examples 
correspond to the normal and self-accelerating branches of pure DGP theory
respectively.} \label{sampleF}

Thus, condition (\ref{F'0}) guarantees that the
Minkowski vacuum is the physical one. However, recall from the
previous section that unless the radion decoupled for one reason or
another, we required $\chi \leq 0$ in order to avoid a ghost. We then
arrive at the following important and somewhat surprising conclusion:
the case of a physically safe, i.e.\ perturbatively stable,  radion mode
corresponds to the unphysical cosmological  case. 
Therefore if we want to stick to a Minkowski vacuum then it follows
that the radion must be absent for the solution to be
consistent. This corresponds to the strongly coupled case ($\chi=0$),
and the cases for which decoupling occurs ($\chi \to \infty$).  We
will study the corresponding cosmological solutions for these
``radion-free'' cases in the next  section.

The link between a consistent cosmology, and the absence of a
perturbative ghost can be traced back to the absence of a tensor
zero-mode. This means that the large distance or late time behaviour is
dominated by the scalar radion (if it exists). Now, if one uses the
Gauss-Codazzi equations to find the projection of the Einstein tensor
on the brane~\cite{projected}, then after linearising, we find  that
\be
\delta \left(\m{R}_\mn-\half
\m{R}\ga_\mn\right)=\frac{1}{2 \chi} T_\mn\label{linG}
\ee
We have left out the explicit contribution from the bulk Weyl tensor
since it is only expected to behave like a dark  source of
radiation~\cite{bwholog}, and is actually zero for the cosmological
branes considered here. Now the effective equation of motion
(\ref{linG}) would follow from an effective action of the  form
\be
S_\tt{eff}=\int  \sqrt{-\ga}\left(\chi \m{R}+\m{L}_\tt{matter}\right) 
\label{s4d}
\ee  
At late times, we now have a natural interpretation for $\chi$: it is
the effective $4D$ Planck scale. Taking this to be positive, it is
clear that we will have  a consistent $4D$ cosmology. However, in the
absence of a tensor zero mode, the conformal mode has nothing to mix
with in the far infra-red, giving rise to a dangerous ``conformal  ghost''.

Note that the condition $F'(0)\geq 0$ is actually required for the 
whole cosmology 
\be
F'(H^2) \geq 0, \qquad \tt{for $0 \leq H < H_\tt{max}$} \label{condF'}
\ee
otherwise the Strong Energy Condition will fail to hold\footnote{
To see this, note that our requirement that $F'$ is positive in a 
neighbourhood of the origin means that if $F'$ becomes negative, then
it must have a zero at finite $H$. This in turn implies that $(\rho+p)=0$,
from (\ref{energy}), thus violating strong energy.}.
The condition (\ref{condF'}) also guarantees that we never enter a
phase of super-inflation, with $\dot H >0$. This can easily be checked
using equation (\ref{rho}) and energy conservation
(\ref{energy}). Super-inflation is usually associated with phantom
cosmologies, which will typically lead to ghost-like
instabilities~\cite{phantom}. 

Last, but not least, we must demand that our cosmology passes through
a standard $4D$ phase at some earlier time, otherwise we will run into
problems with nucleosynthesis~\cite{nucleo}. If $H_\tt{max}$ is
sufficiently large, then when $H \sim H_\tt{max}$, the induced
curvature on the brane will dominate, so that $\rho \sim 6 m_\tt{pl}^2
H^2$. For the accelerating cosmologies to be discussed in the next
section, we will find that $H_\tt{max}$ can be taken to be arbitrarily
large.

\section{Radion-free cosmologies} \label{analysis}

We saw previously  that the only cosmological solutions that were
universally consistent were those for which the radion field was
absent from the vacuum perturbations. This included the conformal
limit ($\chi=0$), and the decoupling limits. The latter corresponds to
two possibilities: a generalised Randall-Sundrum scenario or a
Minkowski bulk on at least one side of the brane. We will look at both
of these cases, as each contains an example of an accelerating cosmology.

\subsection{Acceleration with a decoupled radion} \label{sec:mads}

The generalised Randall-Sundrum scenario 
corresponds to the case where we retain the AdS horizon on either side
of the brane. The scalar excitations about the vacuum decouple. We
also have a consistent cosmology for $0 \leq H < \infty$, defined  by
\be
F(H^2)=6m_\tt{pl}^2H^2+12\langle M^3\left(\sqrt{H^2+k^2}-k\right) \rangle
\ee
One can easily prove that $\m{C}(H^2)=F/F'H^2\geq 1$ for all values of
$H>0$. As discussed in the previous section, acceleration can only
occur for ordinary matter satisfying the strong  energy condition if
$\m{C}(H^2)$ falls below one. We conclude that we can never enter a
phase of cosmic acceleration in the generalised Randall-Sundrum
scenario for ordinary forms of matter. This comes as no surprise,
since the presence of a graviton zero-mode guarantees standard $4D$
behaviour at large distances, and prevents any infra-red modification
of  gravity.

Now consider the case where we have a Minkowski bulk on at least
one side of the brane, and without loss of generality take
$k_2=0$. Again, all scalar excitations about the vacuum
decouple. Having not yet specified the values of $\theta_1$,
$\theta_2$, and $k_1$, the cosmology is, in general, defined  by
\be
F(H^2)=6m_\tt{pl}^2 H^2-6M_1^3 \theta_1 \left( \sqrt{H^2+k_1^2}-k_1
 \right)-6M^3_2\theta_2H \label{Fk2=0}
\ee
If {\it both} sides of the bulk are Minkowski, \ie\ $k_1=0$, then 
\be
F({H}^2)=6 m_{pl}^2 {H}^2- 12 \langle M^3 \theta \rangle {H} 
\ee
this is the DGP model \cite{DGP}.
Whenever $\langle M^3 \theta \rangle=0$, we see that the brane
cosmology receives no contribution from the bulk, and behaves exactly
as for 4 dimensional GR, without a cosmological constant. As is well
known, this will not give us any acceleration. Now consider what
happens when  $\langle M^3 \theta \rangle \neq 0$. For small  ${H}>0$,
we  have
\be
F({H}^2) \sim - 12 \langle M^3 \theta \rangle {H}+\mc{O}({H}^2)
\ee
Since $\rho \geq 0$, we clearly require $ \langle M^3 \theta
\rangle<0$, otherwise the flat vacuum would represent an {\it
isolated} vacuum, as discussed in the previous section. However,
note that when $ \langle M^3 \theta \rangle<0$, we have
$\m{C}(H^2)\geq 1$, for all values of $H>0$, and there is never cosmic
acceleration for ordinary forms of  matter.

The first example of an accelerating cosmology occurs when we have
Minkowski space on one side of the brane, and AdS space on the other.
This corresponds to the case $k_1 \neq k_2=0$, with the cosmology defined by
(\ref{Fk2=0}). Now for small $H>0$, we  have
\be
F(H^2) \sim -6M^3_2\theta_2 H +\m{O}(H^2)
\ee
In order to avoid isolating the flat vacuum, we clearly require that
$\theta_2=-1$.  If, in addition, we take $\theta_1=-1$, it is easy to
check that $\m{C}(H^2) \geq 1$ for all values of $H>0$,  and so there
can never be any acceleration for ordinary forms of matter. Perhaps we
should not be surprised that acceleration is impossible when the warp
factor decays away from the brane on the AdS side, since this will
induce a degree of localisation. In contrast, when the warp factor
grows away from the brane, we might expect more interesting dynamics,
since the graviton will want to localise away from the brane, near the
AdS boundary. This does indeed happen, and, as we shall now
demonstrate in detail,  cosmic acceleration can occur for ordinary
matter if we set $\theta_1=1 $.

In the rest of this subsection, we will focus our attention on the
case where the AdS boundary is included in $\m{M}_1$, and we have
Minkowski space in $\m{M}_2$. The resulting cosmology is defined  by
\be
F(H^2)=6m_\tt{pl}^2 H^2-6M_1^3 \left( \sqrt{H^2+k_1^2}-k_1 \right)+6M^3_2H 
\ee
Now if $M_1\leq M_2$, it is relatively easy to check that $\m{C}(H^2)
\geq 1$ for all $H>0$, and as such, acceleration can never occur for
ordinary matter. We can understand this result as follows. Although
the warp factor grows away from the brane in $\m{M}_1$, the graviton
is not so strongly localised on the AdS boundary, since a small value
of $M_1$ makes it easier for the graviton to propagate towards the
brane. Thus, the degree of delocalisation away from the brane on the
AdS side is not as severe as it might have been for larger values of $M_1$.

Now let us focus in detail on the case where 
the degree of delocalisation is more severe, so that cosmic 
acceleration is maximized, by taking $M_1 > M_2$. It follows that
$F'(H^2)$ has one local minimum in $H>0$, where it takes the  value
\be
F'_\tt{min}=6 m_\tt{pl}^2-3\frac{M_1^3}{k_1}\left(1-\frac{M_2^2}
{M_1^2}\right)^{\frac{3}{2}}
\ee
Recall that the cosmological phase of interest corresponds to $F'(H^2)
>0$ for nonzero $H$. Therefore, if $F_\tt{min}' \leq0$, we 
can only have a consistent cosmology for $0\leq H <H_\tt{max}$ 
where $H_\tt{max}$, is {\it finite}. For $0\leq H <H_\tt{max}$ , 
we can use the fact that
$F''(H^2) \leq 0$ to show that $\m{C}$ is an increasing function, and
so $\m{C}(H^2) \geq \m{C}(0) = 1$, thus cosmic acceleration
can never occur for ordinary matter if $F'_\tt{min} \leq 0$.
In contrast, if $F'_\tt{min} > 0$, then it follows that $F'(H^2)
> 0$ for all $H>0$. There is only one cosmological phase, so we can
take $H_\tt{max}$ to be infinite. This condition on $F'_\tt{min}$
enables us to place the following lower bound on  $m_\tt{pl}^2$
\be
m_\tt{pl}^2  > \frac{M_1^3}{2k_1}\left(1-\frac{M_2^2}
{M_1^2}\right)^{\frac{3}{2}}
\label{mpl}
\ee
Given that our cosmology is consistent for arbitrarily large values of 
$H$, it is instructive to consider the asymptotic behaviour of $\m{C}$:
\be
\m{C} =1-\left[\frac{M_1^3-M_2^3}{m_\tt{pl}^2}\right] H^{-1}+\m{O}(H^{-2})
\ee
There are two things to note here. First, at very large $H$,
$\m{C}\approx 1$ to leading order, so the standard $4D$ cosmology is
reproduced at early times. This is due to the induced curvature term
on the brane dominating the UV behaviour. Second, the first correction
demonstrates that $\m{C}$ starts to fall {\it below} one, since $M_1 >
M_2$, which is precisely the sort of behaviour we hope to see in order
to get cosmic acceleration from ordinary matter. Quite how much
acceleration can be obtained depends on the nature of the cosmological
fluid (the value of $w$), and the minimum value of
$\m{C}(H^2)$. It is the latter that measures the degree to which
modified gravity is contributing to the  acceleration.

To maximize cosmic acceleration, we therefore require the
minimum possible value of $\m{C}(H^2)$:
\be
\m{C}=\mc{C}({\hat H},\beta,\epsilon)
= \frac{(\cos^3\beta + \epsilon) {\hat H}^2 - 2(\sqrt{1 + {\hat H}^2} -1)
+ 2{\hat H} \sin^3 \beta }
{(\cos^3\beta + \epsilon) {\hat H}^2 - {\hat H}^2 / \sqrt{1 + {\hat H}^2}
+ {\hat H} \sin^3 \beta }
\ee
where we have defined
\be
{\hat H} = \frac{H}{k_1}, \qquad \frac{M_2}{M_1}=\sin\beta, 
\qquad \frac{2 m_\tt{pl}^2 k_1}{M_1^3}=\cos^3\beta+\epsilon
\ee
with  $0 \leq \beta < \pi/2$, and $\epsilon > 0$, which is
consistent with  $M_1>M_2$ and the bound (\ref{mpl}).
Unfortunately there is no simple analytic minimization of this
functional, however, it is not difficult to establish that the 
minimum occurs along $\epsilon=0$, i.e.\ on the boundary of our
allowed range of the 4-dimensional Planck mass. Therefore, we can
never actually attain the maximal value of acceleration, however,
numerically, we see that the minimum values one can typically achieve
are of order $\m{C} \sim 0.43$. 
For a matter dominated universe ($w=0$), we can
therefore choose $\beta$ and $\epsilon$ such that the effective
equation of state, $\gamma_\tt{eff}$,  falls as low as $\sim
-0.57$, and for radiation ($w=1/3$) $\sim 0.43$. 

To see if this cosmological model is viable, it is not sufficient
to demonstrate that the effective equation of state bottoms out at some
reasonable negative value, as it may be that the actual cosmological 
evolution does not spend sufficient time in this negative region to
have a significant era of late time acceleration. We therefore need to
track the scale factor throughout time to demonstrate that the period 
and amount of acceleration is at least potentially sufficient.  
Differentiating the Friedmann equation gives the cosmological evolution
of $H$:
\be
\frac{d{\hat H}}{d{\hat t}} = - \frac{3}{2} (1+w){\hat H}^2 \m{C}({\hat H}^2)
\label{Heqn}
\ee
where ${\hat t} = k_1 t$. This is, in general, a somewhat involved 
NLDE for ${\hat H}$, and is best integrated numerically.
We illustrate acceleration in this model by taking a specific example,
for which $\m{C}$ is shown in figure \ref{fig:minkcosmo}.
There are a number of generic features apparent in this example. At
early times, when $H$ is large ($\alpha \approx \pi/2$), $\m{C}$ is
close to one, and so we are in the standard $4D$ regime, because the
induced curvature term dominates in the UV. As the universe starts to
expand, and cool down, then $\m{C}$ starts to fall below one. If we
are in a period of matter domination ($w=0$), we enter an
accelerating phase when $\m{C}$ falls below $2/3$.  The minimum value
of $\m{C}$ is around $0.5$, at which point we have maximal
acceleration, whatever the cosmological fluid. $\m{C}$ then starts to
increase, and we exit the accelerating phase for good. A maximum value
of  $\m{C}$ occurs when ${\hat H} \sim \beta$. Here we are in the phase
of greatest deceleration. This maximum value actually diverges when
$\epsilon=0$, at the point at which $F'$ vanishes. Beyond this phase,
at very late times, $\m{C}$ approaches its limiting value of  $2$,
and hence the cosmology becomes that of a ``stiff matter'' universe,
where $p = \rho$.
\FIGURE{\includegraphics[width=10cm, height=5cm]{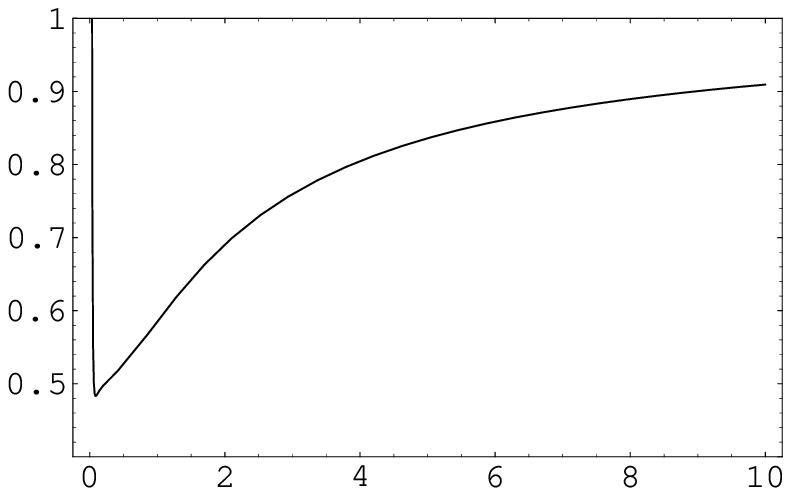}
\caption{Plot of $\m{C}$ against ${\hat H}$, for $\beta=0.02$ and
$\epsilon=0.0002$.}
\label{fig:minkcosmo} }

The evolution of the scale factor can be found by integrating (\ref{Heqn}),
and the comoving Hubble radius is plotted in figure \ref{fig:minkcomov} 
for both matter and radiation dominated cosmologies.
This shows that while the matter cosmology experiences
a significant era of acceleration, the radiation cosmology has
only a period of marginal acceleration, which would be more accurately 
described as `coasting' \cite{coast}.
\FIGURE{\includegraphics[height=4.5cm]{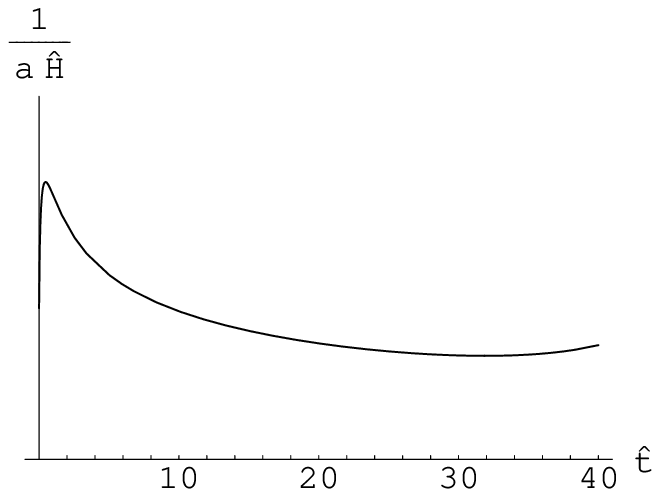}
\nobreak\hskip3mm\nobreak
\includegraphics[height=4.5cm]{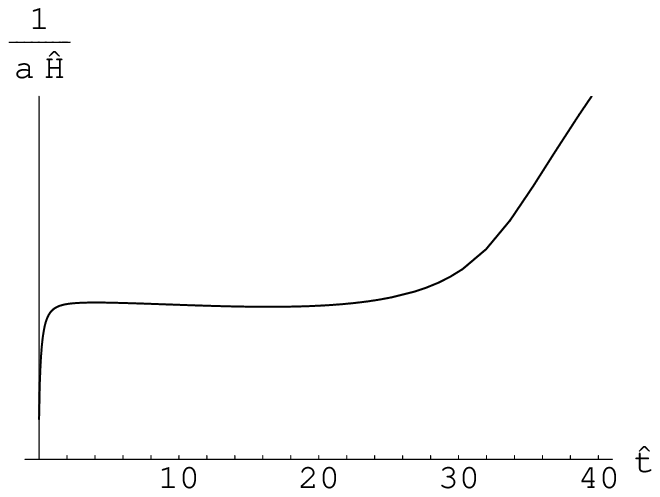}
\caption{Plot of the comoving Hubble radius as a function of
time, measured in units of $k_1^{-1}$, for a) a matter cosmology,
and b) a radiation cosmology, with the same values of $\beta$ 
and $\epsilon$ as in figure \ref{fig:minkcosmo}.}
\label{fig:minkcomov} }

Focussing on the matter cosmology, from
the plot of the comoving Hubble radius, we see that we enter a 
period of acceleration at around $t \simeq 0.5 k_1^{-1}$, and remain in
an accelerating phase until $t\simeq 32 k_1^{-1}$. Since we chose a flat
universe, our value of $k_1$ is determined by the current value of
$\Omega_m $, and the Hubble parameter $H_0$. Since we are using this
model for illustrative purposes, we have not performed a best fit of the
cosmological parameters in our model to the data, however, choosing 
some reasonable values of $k_1$ gives the luminosity-redshift plot
of figure \ref{fig:dlz}. 
\FIGURE{\includegraphics[width=8cm]{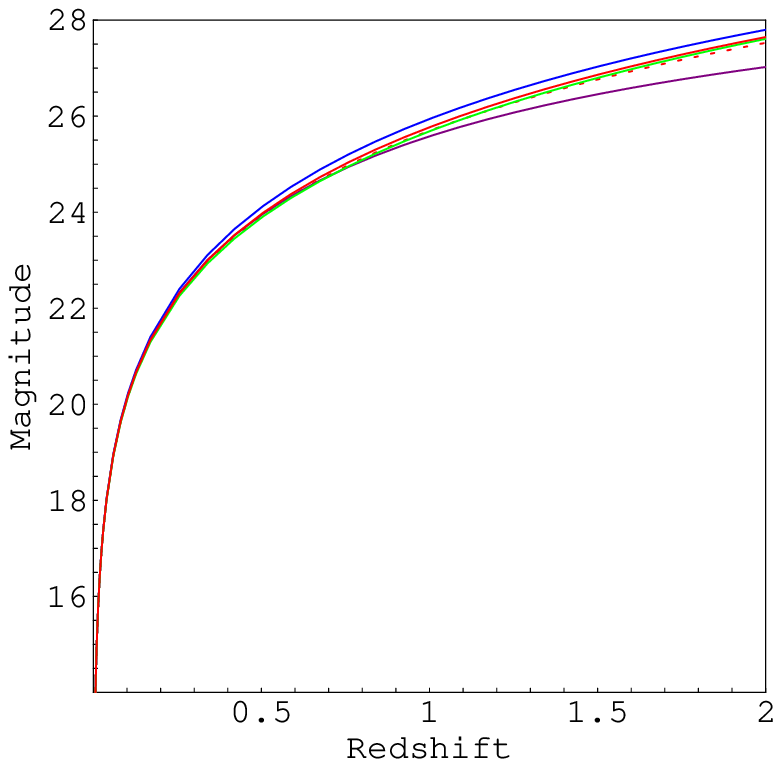}\nobreak\hskip3mm\nobreak
\includegraphics[width=8cm]{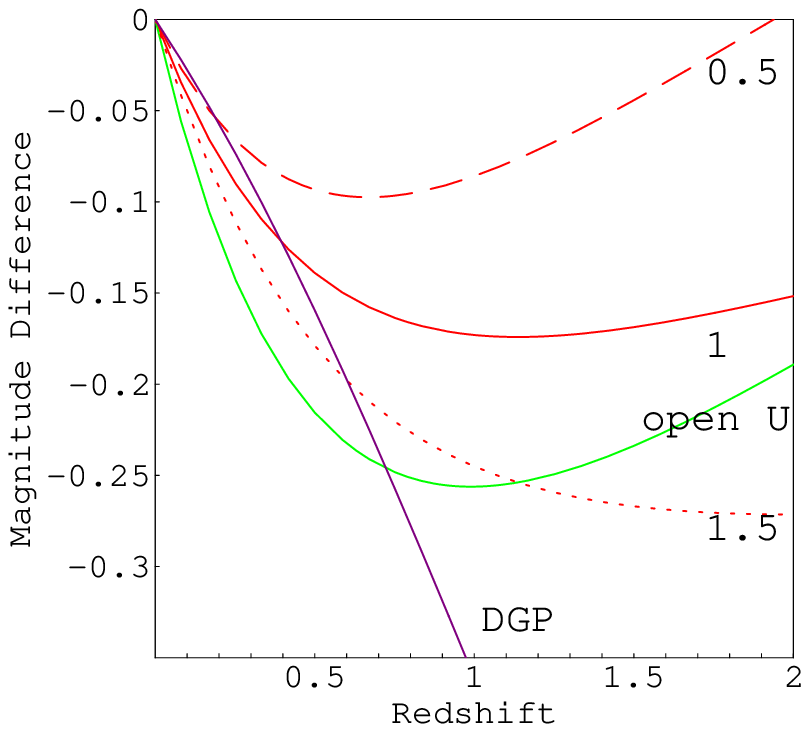}
\caption{Magnitude redshift plots of our cosmology (red curves) for a range
of realistic $k_1$ compared to: $\Lambda$CDM  $(\Omega_m=0.3, 
\Omega_\Lambda=0.7)$  (blue), open matter $(\Omega_m=0.3, \Omega_\Lambda=0)$ 
(green), and the DGP model (purple).  The second plot shows a subtraction
of the predicted magnitudes from $\Lambda$CDM. The labels on the red plots 
show the values of the parameter $k_1$ as a multiple of $H_0$.}
\label{fig:dlz} }

Clearly the model gives a plausible fit
to the data, however, more work needs to be done to establish its
viability, as well as its comparative merits to other models \cite{Liddle}.
In particular, we have taken standard typical values of $H_0$ and 
$\Omega_m$, which of course are the result of a fit of Einstein
evolution to various different data sets. It is possible that using
the evolution predicted by our model, the best fit values of these
parameters may be different, which was the motivation for giving a range
of commonsense values of $k_1$. Even so, it does seem as if our model
gives a better fit than flat DGP, even for the worst case scenario.

\subsection{The conformal cosmology}\label{confcos}
The remaining ``radion-free'' cosmology is interesting,
because it leads to the greatest amount of  acceleration. This is the
conformal case, corresponding  to 
\be
\chi=m_\tt{pl}^2-\left \langle \frac{M^3 \theta}{k}\right \rangle=0
\ee
In this limit, an extra symmetry kicks to linear order in perturbation
theory that eliminates the radion degree of freedom. It also prevents
coupling, at linear order,  to the trace of the energy momentum tensor
on the brane. If a non-zero trace is present, the geometry responds
non-linearly. It is precisely this non-linear effect that gives rise
to the acceleration at late times, as we shall soon  see.

The first thing to note is that the cosmology is consistent for all
values of $H\geq 0$, provided  $m_\tt{pl}^2 >0$ and $F''(0) \geq
0$. At early times the induced curvature on the brane guarantees that
we experience the standard $4D$ evolution. The extra dimensions then
begin to open up and we enter an accelerating phase at later times. To
undertand the late time behaviour, let us approximate equation
(\ref{rho}) by a Taylor expansion near  $H=0$
\be
\label{taylor}
\rho={H}^2F'(0)+\frac{{H}^4}{2} F''(0)+\frac{{H}^6}{6} F'''(0)+
\mathcal{O}\left({H}^8\right)
\ee
where
\be
F'(0)=6\chi,\qquad 
F''(0)= 3\left\langle\frac{M^3\theta}{k^3}\right\rangle,\qquad
F'''(0) = -\frac{9}{2}\left\langle\frac{M^3\theta}{k^5}\right\rangle 
\ee
For $\chi \neq 0$, $\rho$ is linear in $H^2$ to leading order. Cosmic
acceleration in this instance is impossible. It is only as we take the
conformal limit, $\chi \to 0$, that the linear term goes away,  and
acceleration becomes possible. This is because $\rho$ is either
quadratic, or even cubic in $H^2$ at late times. This will lead to
power law acceleration,  since
\be
\rho \propto H^{2n} \implies a(\tau) \sim \tau^{\frac{2n}{3(1+w)}}
\ee
The quadratic dependence ($n=2$) occurs when  $F''(0) > 0$, or equivalently 
\be
\left\langle\frac{M^3\theta}{k^3}\right\rangle >0
\ee
Note that we cannot have $\theta_1=\theta_2=-1$, which comes as no
surprise since this is just the generalised Randall-Sundrum scenario
discussed previously, for which gravity is localised
on the brane. To get an idea  of how much acceleration occurs at late
times, note that for small  $H$, 
$
\m{C}(H^2)=\frac{1}{2}+\m{O}(H^2)
$
and so a pressureless cosmological fluid will lead to an effective  
equation of state, $\gamma_\tt{eff}=-1/2$. 

The cubic dependence ($n=3$) occurs when $F''(0)=0$, or equivalently  
\be
\left\langle\frac{M^3\theta}{k^3}\right\rangle =0
\ee
In order to avoid having an isolated vacuum at $H=0$ we must also
require that $F'''(0)>0$. This scenario leads to even more
acceleration:  for small $H$,  $
\m{C}(H^2)=\frac{1}{3}+\m{O}(H^2)
$
and so a pressureless fluid has an effective equation of state
$\gamma_\tt{eff}=-2/3$ at very late times. Naively, we might have
expected the greatest acceleration to occur when the AdS boundary is
retained on {\it both} sides, but this turns out not to be the case
here. If we assume, without loss of generality, that $M_1>M_2$, then
we must also have $\theta_1=1$, $\theta_2=-1$. This means that in
$\m{M}_1$, we retain the AdS boundary, whereas in $\m{M}_2$ we retain
the  AdS horizon.

We can understand this puzzling result as follows. For these conformal
cosmologies, the scalar radion is strongly coupled and therefore plays
a more important role than the spin 2 graviton mode. While the
graviton likes to localise where the warp factor is greatest, the
opposite is true for the radion. If the radion is dominant, as is the
case here, it is clear that gravity on the brane is weakest when the
radion localises away from the brane, near the AdS horizon. Note that
in the previous case of a Minkowski bulk, the radion was {\it weakly}
coupled, and so the graviton  played the dominant  role. 

For comparison with the Minkowski model, we choose a specific
example with $M_2/M_1 = k_2/k_1 = \sin 0.6$, and show the comoving 
Hubble scale and a similar subtracted luminosity redshift plot in figure
\ref{fig:confplots}. Once again, we have not performed a ``best-fit''
analysis, but simply taken two reasonable values of the model parameters.
\FIGURE{\includegraphics[width=8cm]{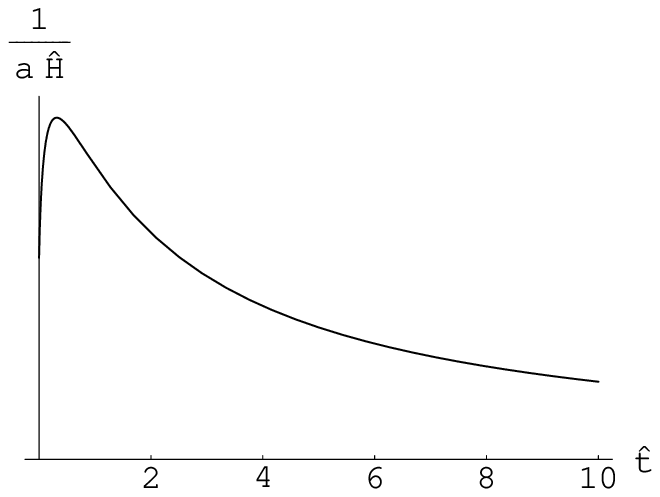}\nobreak\hskip3mm\nobreak
\includegraphics[width=8cm]{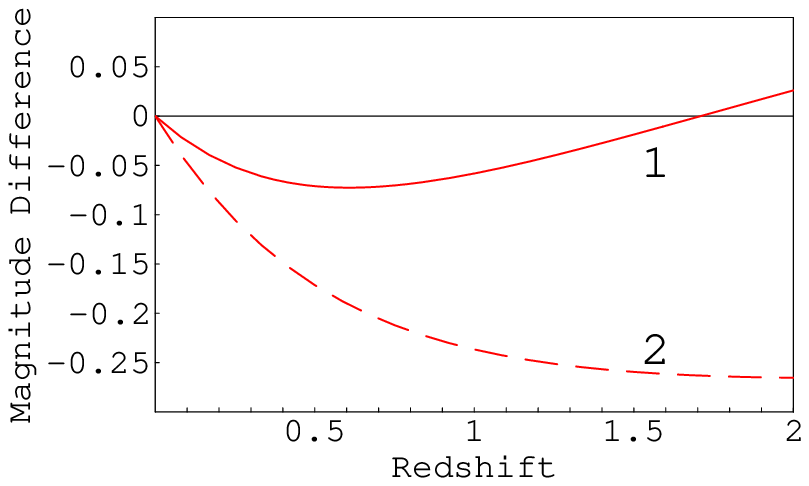}
\caption{The comoving Hubble scale for the conformal cosmology, and the 
subtracted redshift plot (relative to $\Lambda$CDM) for $k_1 = H_0, 2H_0$.
Once again, ${\hat H} = H/k_1$, and ${\hat t} = k_1 t$.}
\label{fig:confplots} }

In principle, if we could make $\rho$ quartic in $H^2$ we could
achieve yet more late time acceleration. 
Unfortunately, this would require us to
set $F'(0)=F''(0)=F'''(0)=0$, which corresponds here to the trivial
case $F(H^2) \equiv 0$. 
One interesting possibility is that the addition of a 
Gauss-Bonnet term would allow this. The Gauss-Bonnet 
term is a higher order curvature correction to the gravitational 
action which is ghost-free (see \cite{fax} for discussions
in the context of braneworlds, and \cite{myers,scd} for the boundary
terms appropriate to Gauss-Bonnet).
We can therefore view it as a possible first correction in a UV
completion of the gravitational effective action. Such UV completions
are sometimes proposed as a remedy to the problems
of the pure DGP model. In this context, the effect of the GB term
is to modify the relation between the constant $k$, and the 
cosmological constants on each side of the brane:
\be
4\alpha_i k_i^2=1\mp\sqrt{1+ 4 \alpha_i \Lambda_i/3}
\ee
where we allow $\alpha$ to take different values on either side of the brane,
in keeping with the asymmetric set-up. 
The brane tension is also modified to: 
\be 
\label{liverpool} \sigma=-12\langle M^3 \theta
k(1-\frac{4}{3}\alpha k^2)\rangle  \; ,
\ee 
The modified
Friedmann equation is again (\ref{rho}) but with F given by 
\be 
F(H^2)= 6m_{pl}^2 H^2-12\left\langle M^3 \theta
\left(\sqrt{H^2+k^2}-k \right)(1-\frac{4}{3}\alpha k^2)\right\rangle 
-32 \left\langle M^3 \theta  \alpha H^2 \sqrt{H^2 +k^2} \right\rangle. 
\ee 
From this, we now see that the Gauss-Bonnet term does not significantly
modify our conclusions, although we cannot push the conformal model to
stronger late time acceleration, since this is incompatible with good
UV behaviour of the cosmology. Thus, we are driven to the interesting 
conclusion that UV corrections in this case appear to dampen acceleration,
although we have not performed a full parameter search to check if there
are regions of the GB theory with greater acceleration.

\section{Discussion} \label{discussion}

Up until now, braneworld models have offered only one new mechanism
for reproducing cosmic acceleration. These are the self-accelerating
de Sitter like solutions found for DGP branes~\cite{DGP} that have
been shown to be unstable at the level of linearised
perturbations~\cite{DGPghosts}. In this paper we have shown that a
dynamical alternative mechanism exists whereby the vacuum brane is
Minkowski, but when matter is present the universe can undergo a
period of acceleration. Quite how much acceleration occurs
depends on the details, but we have demonstrated two types of model
for which there is a convincing amount of acceleration.

The idea of ordinary matter driving accelerated expansion is not new,
for example, the Cardassian model of Freese and Lewis, \cite{Cardassian},
obtained such acceleration by empirically modifying the Friedmann equation,
sending $\rho \to \rho + c \rho^n$, where $n<2/3$ was some 
(unknown) parameter. While an interesting empirical model, Cardassian
expansion did not provide an explanation of why the matter source 
contributed to the Friedmann equation in this way. In our model, we use
concrete five dimensional physics to give rise to 
four dimensional modifications, where it is the geometry which is
altered, rather than the matter source. 
Of course, one could always rewrite the modified Friedmann equation as 
\be
H^2 = \m{F}(\rho)
\ee
by algebraically solving $\rho = F(H^2)$, however, as this involves the
solution of a quartic in either $H$ or $H^2$,  it is both messy, and one
also has to take great care not to pick up fictitious additional 
solutions from branch choices. Nonetheless, one could view our models
as providing a sound theoretical basis for an effective Cardassian model.
For example, the Minkowski-AdS model illustrated in section
\ref{sec:mads} is approximately equivalent to 
Cardassian expansion with $n=0.5$. 

The key feature of our braneworld model for modified gravity is that 
we have shown that our vacuum solutions are perturbatively  stable.
The requirement of perturbative stability is obviously
crucial. Typically, instabilities appear in the scalar sector so we
have focused our attention on scalar perturbations about the Minkowski
brane vacuum. It turns out that there is a curious interplay between
physically acceptable cosmologies, and stable vacua: if the
cosmological solution is well behaved close to the vacuum, then the
radion mode is a ghost (and vice-versa!) This surprising result can be
understood as stemming from the ``conformal ghost'', as discussed in
detail at the end of section \ref{pert}. Perhaps more importantly, it
greatly restricts the set of physically acceptable set-ups: we can
only consider those cases for which the radion either decouples
completely ($\chi \to \infty$), or is eliminated by some extra
symmetry as in the conformal case  ($\chi=0$).

The conformal cosmology discussed in section \ref{confcos} leads to
the greatest amount of acceleration. Although we prefer to regard it
as a stand alone special case of enhanced symmetry, we can gain some
understanding of its dynamics by considering the limit of small $\chi
\neq 0$. The first thing to note is that the radion coupling scales
like $1/\chi$, so we might also think of the conformal case as the
strongly coupled limit. Because the radion is so strongly coupled it
dominates the dynamics over the the spin 2 mode. Therefore the case of
greatest acceleration occurs when one has the adS boundary   on one
side of the brane, and the adS horizon on the other. The presence of
the adS boundary ensures that the radion does not decouple,  and is
able to  draw gravity away from the brane by localising close to the
adS  horizon!  

The case of $\chi$ small and positive may also be of
phenomenological interest. The point is that although our linearised
analysis suggests that the radion is a ghost, it is so strongly
coupled that one cannot fully trust linearised perturbation theory
except on the largest scales. One could therefore imagine a scenario
whereby that scale is pushed beyond (say) the size of the observable
horizon. Our cosmological solution is a non-linear solution so  we can
retain the power law acceleration in the presence of matter or  radiation. 

Naturally, our models, while appearing to give a promising fit to the
SN data, are not necessarily preferable to $\La$CDM. The main problem
is the extension of parameter space - we can explain late time
acceleration, but at the cost of an additional parameter. If we suppose
that our four dimensional Planck scale is given, then our models have
in general three free parameters\footnote{Note that we are referring 
here to the parametrization of the matter or geometry, rather than
the whole gamut of cosmological parameters. }: $k_1$, $M_1$ 
and $M_2$ (however, note
that the conformal cosmology has only {\it two} free parameters, $k_1$ and
$M_1$, say). This is to be compared to the two parameters of $\La$CDM,
$\Omega_m$ and $\Omega_\La$. Clearly then, our models will in general fail on
selection criteria (see e.g.\ \cite{AIC} for a discussion
of this issue in the current context), unless significant 
deviations to $\La$CDM are found at higher redshift.
However, our conformal cosmology, for which the parameter space is not
higher dimensional, certainly merits further investigation.

In this paper, we have been primarily concerned with producing a
{\it consistent cosmology}, i.e.\ a model which has consistent 
particle physics (such as no ghosts) and also provides an accelerating
cosmology. One question we have not considered is whether table-top or 
indeed solar system / astrophysics is consistent. The DGP model has
received a great deal of attention, mainly because it appeared
to provide a ghost-free way of getting the gravitational phenomenology
first discovered in the GRS model, however, more careful analysis
uncovered a number of difficulties (such as strong coupling) and
a fundamental flaw: the existence of ghosts. Here, we have resolved
the fundamental flaw by introducing an asymmetric braneworld, but we have
not provided a solution (or indeed an examination) of any of the 
difficulties. For example, strong coupling, which is an issue 
in the DGP model, is also an issue in our conformal model. 

Asymmetric braneworlds are an under-explored territory in the world of
brane model building, largely because the asymmetry introduces additional
complications at the level of the equations of motion. The main 
rationale for considering symmetric braneworlds is that these
correspond to orbifold compactifications in string theory. However, 
the string landscape is also a popular concept \cite{landscape}, 
and in the landscape we can
expect many different vacua, with many different local cosmological 
constants \cite{BP}, and indeed (through warping \cite{GKP}) the 
possibility of different 
local effective Planck masses. Therefore, although the idea of having
different Planck masses and cosmological constants on either side of 
the brane may seem a little strange at first sight, it is in fact
a natural, if not generic, occurance in the context of the string
landscape.

We end by re-emphasizing that we do not intend to present this as an
improvement on the $\La$CDM model. We merely wish to demonstrate what
can be achieved consistently in a braneworld set-up. 
Ultimately, the concordance model is preferable because it provides
the {\it simplest} fit to the data - i.e.\ the fit with the smallest number
of parameters. However, the definition of `simplest' should also include
a measure of how naturally the model fits within an underlying fundamental
theory of interactions. Whether it is easier to produce a massively fine
tuned cosmological constant, or discrete vacua with differing parameters
(or something else), will no doubt continue to the a source of lively 
debate for some time!


~\\
{\large \bf Acknowledgements}
We would like to thank Ed Copeland, Damien Easson, Nemanja Kaloper,
Gustavo Niz and Tom Shanks for useful discussions.

\end{document}